\documentclass{jnmp03c}

\begin{document}
\setcounter{page}{96}

\renewcommand{\evenhead}{J Chavarriga and  I A Garc\'{\i}a}
\renewcommand{\oddhead}{Integrability and Explicit Solutions
in Some Bianchi Systems}

\thispagestyle{empty}

\FistPageHead{1}{\pageref{chavarriga-firstpage}--\pageref{chavarriga-lastpage}}{Article}

\copyrightnote{2001}{J Chavarriga and  I A Garc\'{\i}a}

\Name{Integrability and Explicit Solutions \\
in Some Bianchi Cosmological Dynamical Systems}\label{chavarriga-firstpage}

\Author{J CHAVARRIGA~$^1$ and I A GARC\'{I}A~$^2$}

\Adress{Departament de Matem\`atica, Universitat de Lleida\\
Avda. Jaume II, 69, 25001, Lleida, Spain \\
$^1$~E-mail: chava@eup.udl.es\\
$^2$~E-mail: garcia@eup.udl.es}

\Date{Received April 6, 2000; Revised August 2, 2000; Accepted September 18, 2000}

\begin{abstract}
\noindent
The Einstein field equations for several cosmological models
reduce to polynomial systems of ordinary differential equations.
In this paper we shall concentrate our attention to the spatially
homogeneous diagonal $G_2$ cosmologies. By using Darboux's
 theory in order to study ordinary differential equations in
the complex projective plane~${\mathbb C}{\mathbb P}^2$ we solve
the Bianchi~V models totally. Moreover, we carry out a study of
Bianchi~VI models and first integrals are given in particular
cases.
\end{abstract}

\section{Introduction}

The Einstein field equations for a number of classes of
cosmological models have been written as polynomial systems of
ordinary differential equations. In recent investigations some
authors look for new exact solutions of these differential systems
due to their cosmological interest. Most of these solutions araise
either due to the existence of a symmetry, or under a restriction
on the parameters of the cosmological model. Homogeneous
cosmological models have important applications in the theory of
evolution of the universe. In particular they can be applied to
the theory of explosions of stars, formation of ga\-la\-xies, etc.
The equations
 of general relativity are reduced to systems of a finite (but quite large)
number of ordinary differential equations. In the last two decades these
dynamical systems were and still are being analyzed by means of traditionals
analytic and numerical methods. However the dynamical systems in question
are so complex and the development of effective methods for their analysis
becomes especially urgent.

In this paper we shall concentrate our attention to the
diagonal $G_2$ cosmologies, i.e. those spacetimes which
admit an Abelian $G_2$ comprising two hypersurface-orthogonal
Killing vector fields (see for instance~\cite{He1}).
In fact we only study spatially homogeneous diagonal $G_2$ cosmologies.
 It is assumed that the Einstein field equations are satisfied
and that spacetime admits a perfect fluid with linear equation of state
\begin{equation}
p=(\gamma - 1) \rho, \qquad \frac{2}{3} < \gamma < 2,
\label{bi11}
\end{equation}
where $\gamma$ is an arbitrary parameter in the specified
range of values which contains the physically significant values
$\gamma = 1$ (corresponding to dust) and $\gamma = 4/3$
(for which the fluid is interpreted as radiation). In all the paper
we use geometrized units with $c=1$ and $8 \pi G=1$ and the
sign convention of MacCallum~\cite{Ma}.

In Section~2 we describe the mathematical techniques used for
the analysis of polynomial differential equations that we deal in this work.
A summary is given about the Darboux's theory in order to study ordinary
differential equations in the complex projective plane
${\mathbb C}{\mathbb P}^2$ and, in short, its utility for the analysis
of the socalled {\it degenerate infinity} systems.
This name is due to the fact that in the Poincar\'e compactification
of these kind of differential systems all the equator of the sphere $S^2$,
i.e. the infinity of the affine plane, is filled up of critical points,
see Sotomayor~\cite{Sotomayor2}.
In Section~3 we show the equations that govern the spatially
homogeneous diagonal $G_2$ cosmologies and the Bianchi models.
In Section~4 we solve the Bianchi~V
models totally and, finally in the fifth and sixth sections we carry out a study
of the Bianchi VI cosmological models. First integrals are given in
particular cases and a canonical form of the equations allows
us to obtain the formal solution.

\section{Polynomial foliations in $\pbf{{\mathbb C}{\mathbb P}^2}$}

In this section we consider the first order differential equation
\begin{equation}
Q(x,y) \, dx - P(x,y) \, dy = 0,
\label{bi21}
\end{equation}
where $P$ and $Q$ are polynomials which in the following we can
suppose with coefficients in the complex field ${\mathbb C}$. We
say that $m=\max \{ \deg P, \deg  Q \}$ is the {\it degree} of
polynomial differential equation (\ref{bi21}). When we consider it
necessary, we introduce a ``time" parameter~$t$, which enables us
to express (\ref{bi21}) as a first order autonomous system in the
affine plane
\begin{equation}
\dot{x}=\frac{d x}{d t}=P(x,y) , \qquad \dot{y}=\frac{d y}{d t}=Q(x,y).
\label{bi21.1}
\end{equation}

Following Darboux \cite{D}, we introduce the homogeneous coordinates
$(X,Y,Z)$ of the complex projective plane ${\mathbb C}{\mathbb P}^2$ such that
\begin{equation}
x=\frac{X}{Z} , \qquad y=\frac{Y}{Z},
\label{bi22}
\end{equation}
which allow us to express differential equation (\ref{bi21}) in the form
\begin{equation}
L(X,Y,Z) (Y \, dZ- Z \, dY) + M(X,Y,Z) (Z \ ,dX-X \, dZ)= 0,
\label{bi23}
\end{equation}
where $L(X,Y,Z)=Z^m P( X/Z , Y/Z)$
and $M(X,Y,Z)=Z^m Q( X/Z , Y/Z)$ are homogeneous polynomials of degree~$m$.
Note that $Z=0$ is an invariant straight line of this equation.

More generally (see \cite{D}), we can consider a differential
equation defined in ${\mathbb C}{\mathbb P}^2$ as an equation of the form
\begin{equation}
L (Y \, dZ- Z \, dY) + M (Z \, dX-X \, dZ)+ N (X \, dY-Y \, dX)= 0,
\label{bi24}
\end{equation}
where $L(X,Y,Z)$, $M(X,Y,Z)$ and $N(X,Y,Z)$
are homogeneous polynomials such that $\deg L= \deg M = \deg N =m$.
Introducing the auxiliary independent variable $T$, we can also interpret
equation~(\ref{bi24}) like a dynamical system
\begin{equation}
X'=\frac{d X}{d T}= L(X,Y,Z) , \quad Y'=\frac{d Y}{d T}= M(X,Y,Z) ,
\quad  Z'=\frac{d Z}{d T}= N(X,Y,Z),
\label{bi24.1}
\end{equation}
over ${\mathbb C}^3 \backslash \{ 0 \}$.
Such homogeneous system induces on ${\mathbb C}{\mathbb P}^2$
the polynomial foliation given by~(\ref{bi24}).
In fact, differential equation~(\ref{bi24}) has not a unique expression.

\medskip

\noindent
{\bf Remark 1.} If we take $\tilde{L}=L- \Lambda X$,
$\tilde{M}=M- \Lambda Y$ and $\tilde{N}=N-\Lambda Z$
 instead of $L$, $M$ and $N$  respectively into equation (\ref{bi24}),
where $\Lambda(X,Y,Z)$ is a homogeneous polynomial of degree $m-1$,
then differential equation {\rm (\ref{bi24})} remains invariant.

\medskip

\noindent
{\bf Remark 2.} One of the most obvious questions to ask is
which is the relationship between the parameters $t$ and $T$
of the affine differential system (\ref{bi21.1}) and
the projective differential system {\rm (\ref{bi24.1})} with
$N \equiv 0$. A straightforward differentiation shows that
\[
\frac{d x}{d T} = Z^{m-1} \left[ L(x,y,1)- x N(x,y,1) \right],
\qquad \frac{d y}{d T} =Z^{m-1} \left[ M(x,y,1)- y N(x,y,1) \right].
\]
On the other hand, by the chain's rule we have
\[
\frac{d x}{d T} = P(x,y) \frac{d t}{d T},
 \qquad \frac{d y}{d T} = Q(x,y) \frac{d t}{d T}.
\]
Finally, identifying the last two previous expressions when $N \equiv 0$
we conclude that
\begin{equation}
\frac{d t}{d T} = Z^{m-1}.
\label{bi24.2}
\end{equation}

A polynomial differential equation (\ref{bi21}),
 where $P(x,y)= \sum\limits_{k=0}^m P_k(x,y)$
and $Q(x,y)= \sum\limits_{k=0}^m Q_k(x,y)$ is the developement in
homogeneous components of the polynomials $P$ and~$Q$ is said to
be {\it degenerate infinity} if it verifies the condition  $x
Q_m(x,y) - y P_m(x,y) \equiv 0$ or equivalently $P_m(x,y)=x
\Lambda(x,y)$ and $Q_m(x,y)=y \Lambda(x,y)$ for some homogeneous
polynomial $\Lambda(x,y)$ of degree $m-1$. In order to clarify
this terminology, we write equation~(\ref{bi21}) in the complex
projective plane as equation (\ref{bi24}), where
\[
\ba{l}
\ds L(X,Y,Z)  =  \sum_{k=0}^{m-1} P_k(X,Y) Z^{m-k} + \Lambda(X,Y) X ,
\vspace{3mm}\\
\ds M(X,Y,Z)  =  \sum_{k=0}^{m-1} Q_k(X,Y) Z^{m-k} + \Lambda(X,Y) Y ,
\vspace{3mm} \\
N(X,Y,Z)  =  0 .
\ea
\]
Now, from Remark 1, we can take
$\tilde{L}=L- \Lambda X$, $\tilde{M}=M- \Lambda Y$ and
$\tilde{N}=N-\Lambda Z$ obtaining
\begin{equation}
\tilde{L} (Y  \, dZ- Z\, dY) + \tilde{M} (Z \, dX-X \, dZ)+ \tilde{N} (X  \,dY-Y \,dX)= 0 ,
\label{bi25}
\end{equation}
and being
\[
\ba{l}
\ds \tilde{L}(X,Y,Z)  =  Z \sum_{k=0}^{m-1} P_k(X,Y) Z^{m-k-1} ,
\vspace{3mm}\\
\ds \tilde{M}(X,Y,Z)  =  Z \sum_{k=0}^{m-1} Q_k(X,Y) Z^{m-k-1},
\vspace{3mm}\\
\tilde{N}(X,Y,Z)  =  -Z \Lambda.
\ea
\]
From the above equation we conclude that the line of infinity $Z=0$
 is obviously an invariant straight line of (\ref{bi25})
and moreover this line is filled up in a dense sense of critical points
since $\tilde{L}(X,Y,0)= \tilde{M}(X,Y,0) = \tilde{N}(X,Y,0) = 0$.

At this point we want to comment that it is more convenient,
in order to obtain their critical points, write equation (\ref{bi24}) into the form
\begin{equation}
{\cal P} \, dX + {\cal Q} \, dY + {\cal R} \, dZ = 0,
\label{bi26}
\end{equation}
being ${\cal P}=M Z-N Y$,
${\cal Q}=N X- L Z$ and ${\cal R}=L Y-M X$.
We say that $(X_0, Y_0, Z_0) \in {\mathbb C}{\mathbb P}^2$
is a {\it critical point} of equation (\ref{bi26}) if and only
 if ${\cal P}(X_0, Y_0, Z_0)={\cal Q}(X_0, Y_0, Z_0)= {\cal R}(X_0, Y_0, Z_0)=0$.

Finally, we recall that the complex projective plane
${\mathbb C}{\mathbb P}^2$ is compact because it is
obtained by adding a line at infinity to the plane ${\mathbb C}^2$.
 Hence, when we extend the foliations from the affine plane
to the projective plane and we study their integrability, this
study is global in the above sense.

\section{Spatially homogeneous diagonal $\pbf{G_2}$ cosmologies}

Homogeneous cosmological models are defined to be
space-time manifolds thogether with a metric $d s^2$ satisfying
Einstein's equations and invariant under some three-dimensional
Lie group of transformations of the manifold. In cosmology the
most important models are spatially homogeneous models for which
 the orbits of the Lie group are space-like (restrictions of the metric
$d s^2$ to the orbits of the group are negative definite).

Let $\Sigma_+$ and $\Sigma_-$ be dimensionless shear components.
In addition, the variables $A$ and $N_\times$ are dimensionless
measures of the isotropic and anisotropic curvature of the hypersurfaces
which lie orthogonal to the fluid flow. It is easy to show (see \cite{He1}
and \cite{He2} for example) that the Einstein field equations for the
spatially homogeneous diagonal $G_2$ cosmologies reduce to the
following system of polynomial ordinary differential equations
\begin{equation}
\ba{ll}
\dot{\Sigma}_+ = (q-2) \Sigma_+ -2 N_\times^2 ,\quad &
\dot{\Sigma}_- = (q-2) \Sigma_- +2 A N_\times,
\vspace{2mm}\\
\dot{N}_\times = (q+2 \Sigma_+) N_\times , & \dot{A} = (q+2 \Sigma_+) A ,
\ea
\label{bi41}
\end{equation}
where the over dot denotes $d / d \tau$, i.e.,
derivative with respect to a dimensionlees measure of time
$\tau$ along the fluid flow lines (see \cite{Wa}). Moreover,
the deceleration parameter $q$ is given by
\begin{equation}
q= 2 \left(\Sigma_+^2+ \Sigma_-^2\right) + \frac 12
(3 \gamma -2) \Omega,
\label{bi42}
\end{equation}
and the density parameter adopts the form
\begin{equation}
\Omega= 1-\Sigma_+^2- \Sigma_-^2-A^2-N_\times^2.
\label{bi43}
\end{equation}

For this cosmological model it is known that the
Einstein field equations (\ref{bi41}) admit the particular solution
\begin{equation}
 A \Sigma_+ + N_\times \Sigma_- = 0.
\label{bi44}
\end{equation}

The constraint equation (\ref{bi44})
allows the following possibilities (each one of them gives rise to a Bianchi model).
\begin{itemize}
\topsep0mm
\partopsep0mm
\parsep0mm
\itemsep0mm
\item If $A=N_\times =0$ we have the model {\bf B(I)}.
\item Imposing $\Sigma_+=N_\times =0$ the resultant model is {\bf B(V)}.
\item From the last two equations of (\ref{bi41}) it is obvious that
if $N_\times \neq 0$, then system (\ref{bi41}) admits the first integral
$A=k N_\times$ being $k$ a constant. Therefore,
constraint (\ref{bi44}) implies $\Sigma_-=-k \Sigma_+$
and the cosmological model is called {\bf B(VI)}.
\end{itemize}

The Bianchi I models {\bf B(I)} are expressed by the following system
\begin{equation}
\dot{\Sigma}_+ = (q-2) \Sigma_+, \qquad \dot{\Sigma}_- = (q-2) \Sigma_-,
\label{bi45}
\end{equation}
where $q$ is given by (\ref{bi42}) and $\Omega= 1-\Sigma_+^2- \Sigma_-^2$.
These models are easily integrated to yield either $\Sigma_-=k \Sigma_+$,
where $k$ is an arbitrary constant or $\Sigma_+=0$.
Therefore, in the phase portrait we only have singular points and straight lines.

In \cite{Co} the author carries out a qualitative study
of all these Bianchi type models. On the other hand, \cite{He2}
examine the different Bianchi types and determine all of the linear
and quadratic algebraic invariant curves which lie in the physical region
of phase space.

\section{The solution of Bianchi V models}

The Bianchi V models {\bf B(V)}
are given by the following cubic polynomial systems
\begin{equation}
\dot{\Sigma}_- = (q-2) \Sigma_- , \qquad \dot{A} = q A,
\label{bi51}
\end{equation}
where the deceleration $q$ and the density $\Omega$ are
\begin{equation}
q= 2 \Sigma_-^2 + \frac 12 (3 \gamma -2) \Omega, \qquad
 \Omega= 1-A^2- \Sigma_-^2.
\label{bi52}
\end{equation}

These models have been studied by many authors.
In particular, it has been pointed out (see \cite{El,Kr} and \cite{Na})
that the Einstein field equations reduce to elliptic integrals for certain
values of $\gamma$. Moreover, in \cite{He2}
the following time invariant for $\Sigma_- \neq 0$ is given
\[
A= Q_0 \Sigma_- {\rm e}^{2 \tau},
\]
where $Q_0$ is an integration constant.

Performing the change of variables $\Sigma_- = \sqrt{x}$,
$A= \sqrt{y}$ we can reduce the degree of system (\ref{bi51})
obtaining the following quadratic system with degenerate infinity
\begin{equation}
\dot{x} = 3 (\gamma -2) x + x \Lambda(x,y) , \qquad
\dot{y} = (3 \gamma -2) y + y \Lambda(x,y),
\label{B2}
\end{equation}
where $\Lambda(x,y)=3 (2-\gamma) x + (2-3 \gamma) y$.
Applying the arguments of second section, we write system (\ref{B2})
 in the complex projective plane as
\begin{equation}
\tilde{L} (Y \, dZ- Z \, dY) + \tilde{M} (Z \, dX-X \, dZ)+ \tilde{N} (X \, dY-Y \, dX)= 0,
\label{B2.1}
\end{equation}
where, making the rescaling of time $d \tau = Z\, dT$ (see Remark 2), we have
\[
\tilde{L}= 3 (\gamma -2) X , \qquad
 \tilde{M}=(3 \gamma -2) Y , \qquad
 \tilde{N}=-\Lambda(X,Y).
\]

From the first two equations $\tilde{L}$ and $\tilde{M}$
we obtain for equation (\ref{B2.1}) the following first integral
\[
H_1(X,Y,Z)= X^{3 \gamma -2} Y^{3(2- \gamma)} ,
\]
which is a fourth degree homogeneous function. Moreover, as
$\tilde{L}$, $\tilde{M}$ and $\tilde{N}$ are linear homogeneous
polynomials which only depend on the variables $X$ and $Y$, there
exist constants~$c_1$,~$c_2$ and $c_3$ not all zero such that $c_1
\tilde{L}+c_2 \tilde{M}+c_3 \tilde{N} \equiv 0$. Hence, we have
for equation~(\ref{B2.1}) the first integral $H_2(X,Y,Z)=c_1 X+c_2
Y+c_3 Z$. In short we have for instance the linear first integral
$H_2(X,Y,Z)=X+ Y- Z$. Now, we consider for equation (\ref{B2.1})
the zero degree homogeneous first integral defined like
$H_1(X,Y,Z) / H_2^4(X,Y,Z)$. Making the transformation $(X,Y,Z)
\to (x,y,1)$ in the above first integral we turn to the affine
plane obtaining the first integral for system (\ref{B2})
\[
H(x,y):= \frac{H_1(x,y,1)}{H_2^4 (x,y,1)}=
\frac{x^{3 \gamma-2} y^{3 (2- \gamma)}}{(x+y-1)^4}.
\]
Finally, undoing the change of variables $\Sigma_- = \sqrt{x}$,
$A= \sqrt{y}$ we have for Bianchi~V models~(\ref{bi51}) the
following Darboux first integral
\begin{equation}
H(\Sigma_- , A)=
\frac{\Sigma_-^{3 \gamma-2} A^{3 (2- \gamma)}}{\left(\Sigma_-^2+A^2-1\right)^2}.
\label{B2.2}
\end{equation}

Moreover, from the equations of $\tilde{L}$ and $\tilde{M}$
 we obtain the time parametrization
\[
X( \tau )=X_0 \, {\rm e}^{3 (\gamma -2) \tau} ,
\qquad Y( \tau )=Y_0 \, {\rm e}^{(3 \gamma-2) \tau},
\]
where $X_0$ and $Y_0$ are real parameters. In addition,
since $\tilde{N}=-\Lambda(X,Y)$, we can integrate this
equation and write the time parametrization of the variable $Z$ as follows
\[
\ba{l}
\ds Z( \tau )   =  - \int \Lambda(X( \tau ) , Y( \tau )) \, d \tau =
- \int \left[ 3 (2-\gamma) X( \tau ) + (2-3 \gamma) Y( \tau ) \right] \, d \tau
\vspace{2mm}\\
\ds  \phantom{Z(\tau)}
=  - \int \left[ 3 (2-\gamma) X_0 \, {\rm e}^{3 (\gamma -2) \tau}
+ (2-3 \gamma) Y_0 \ {\rm e}^{(3 \gamma-2) \tau} \right] \, d \tau
= X( \tau )+Y( \tau )+C,
\ea
\]
being $C$ a real parameter also. Finally, we turn to the affine plane
 ${\mathbb R}^2$ by means of
$x( \tau )=X( \tau ) / Z( \tau )$, $y( \tau )=Y( \tau ) / Z( \tau )$,
and we have the general explicit solution of Bianchi~V
models (\ref{B2}) given by $\Sigma_- (\tau)= \sqrt{x(\tau)}$,
$A (\tau)= \sqrt{y(\tau)}$. In a more explicit way, taking the initial values
$\Sigma_- (0):= \Sigma_-^0 = \sqrt{X_0 / (X_0+Y_0+C)}$
and $A(0):= A^0 = \sqrt{Y_0 / (X_0+Y_0+C)}$, we have
\begin{equation}
\ba{l}
\ds \Sigma_- (\tau)  =  \Sigma_-^0 \sqrt{
\frac{ \exp( 3( \gamma -2) \tau )}{\left(\Sigma_-^0\right)^2
\left[ \exp( 3( \gamma -2) \tau ) - 1 \right] + (A^0)^2
[ \exp( (3 \gamma -2) \tau ) -1 ] + 1}} ,
\vspace{4mm}\\
\ds A( \tau )  = A^0 \sqrt{\frac{\exp( (3 \gamma -2) \tau )}{\left(\Sigma_-^0\right)^2
[ \exp( 3( \gamma -2) \tau ) - 1 ] + (A^0)^2 [ \exp( (3 \gamma -2) \tau ) -1 ] + 1}}.
\ea\label{solbiV}
\end{equation}

\section{Bianchi VI models}

The Bianchi VI models {\bf B(VI)} are given by the following cubic polynomial system
\begin{equation}
\dot{\Sigma}_+ = (q-2) \Sigma_+ -2 N_\times^2 , \qquad
 \dot{N}_\times = (q+ 2 \Sigma_+) N_\times,
\label{biVI1}
\end{equation}
where the deceleration $q$ and the density $\Omega$ are
\begin{equation}
q= 2 \left(1+k^2\right) \Sigma_+^2 + \frac 12 (3 \gamma -2) \Omega,
\qquad \Omega= 1-\left(1+k^2\right)\left( N_\times^2 + \Sigma_+^2 \right).
\label{biVI2}
\end{equation}
These models admit the invariant algebraic curves $N_\times =0$
 (Bianchi~I solutions)
and $\Omega = 0$ (Ellis--MacCallum vacuum solutions, see~\cite{Kr}
p.~136).

In \cite{He2} Hewitt shows that in the particular case
\begin{equation}
1+k^2= \left( \frac{8}{3 \gamma + 2} \right)^2,
\label{biVI3}
\end{equation}
system (\ref{biVI1}) possesses another invariant algebraic curve given by
\begin{equation}
\Delta:= \left( \Sigma_+ +\frac{3 \gamma +2}{16} \right)^2
+\frac{N_\times^2}{2}-\left( \frac{3 \gamma +2}{16} \right)^2 = 0.
\label{biVI4}
\end{equation}
Furthermore, for this value of the parameter $k$,
he shows that system (\ref{biVI1}) has the time invariant
\begin{equation}
N_\times^{6(2-\gamma)} \Delta^{3 \gamma +2} \Omega^{-8} =
Q_0 \exp\left( -\frac{9}{2}(2-\gamma)^2 \tau \right).
\label{biVI5}
\end{equation}

But, with these three invariant algebraic curves, we can be able
of construct for Bianchi~VI models (\ref{biVI1})
 with restriction (\ref{biVI3}) the integrating factor
\begin{equation}
R(\Sigma_+ , N_\times):=
 N_\times^{-3} \Delta^{\frac{3 \gamma +2}{3 (\gamma -2)}}
 \Omega^{\frac{9 \gamma -2}{6(2-\gamma)}}.
\label{biVI6}
\end{equation}

There exists another particular case of Bianchi VI model given by the constraint
\begin{equation}
1+k^2= \left( \frac{2}{3 \gamma -4} \right)^2,
\label{biVI7}
\end{equation}
for which system (\ref{biVI1}) admits the
following new quadratic invariant algebraic curve
\begin{equation}
\Gamma:= (2-3 \gamma) (3 \gamma-4)^2
+4 (4-3 \gamma)(3 \gamma-2) \Sigma_+
+ 4 (2-3 \gamma) \Sigma_+^2 +12 (2-\gamma) N_\times^2 = 0.
\label{biVI8}
\end{equation}
This invariant curve allows us to obtain for Bianchi VI
models (\ref{biVI1}) with restriction (\ref{biVI7}) an integrating factor given by
\begin{equation}
R(\Sigma_+ , N_\times):=
N_\times^{\frac{2(4-3 \gamma)}{3 \gamma -2}}
\Gamma^{{3 \gamma -4}{3 \gamma -2}} \Omega^{-2}.
\label{biVI9}
\end{equation}

Thus, the first integral $H$ associated to the integrating factor $R$ is given by
\[
H(\Sigma_+ , N_\times)= \int \left[ (q-2) \Sigma_+ -2 N_\times^2 \right]
R(\Sigma_+ , N_\times) \, d N_\times + F( \Sigma_+),
\]
where the function $F$ is calculated as usual by the condition
\[
\frac{\partial H}{\partial \Sigma_+}=
- R(\Sigma_+ , N_\times) (q+ 2 \Sigma_+) N_\times.
\]

Another point of view in order to study Bianchi VI cosmological
models it is the transformation of system (\ref{biVI1})
to the projective differential equation (\ref{bi23}) by means
of the change to homogeneous coordinates $\Sigma_+ = X / Z$,
$N_\times = Y / Z$. Thus, after the application of Remark~1
we obtain that system (\ref{biVI1}) adopts the form
\begin{equation}
\tilde{L} (Y \, dZ- Z \, dY) + \tilde{M} (Z \, dX-X \, dZ)+ \tilde{N} (X \, dY-Y \, dX)= 0,
\label{biVI10}
\end{equation}
where
\[
\ba{l}
\ds \tilde{L}(X,Y,Z)  =  \frac 12  Z \left( 3(\gamma-2) X Z-4 Y^2 \right),
\vspace{3mm}\\
\ds \tilde{M}(X,Y,Z)  = \frac{1}{2} Z Y \left( 4 X+(3 \gamma-2) Z \right),
\vspace{3mm}\\
\ds \tilde{N}(X,Y,Z)  =  -Z \Lambda(X,Y,Z),
\ea
\]
and being $\Lambda(X,Y,Z):= \frac 12 \left(1+k^2\right)
\left[ 3(2- \gamma) X^2+(2-3 \gamma) Y^2 \right]$.
Differential equation (\ref{biVI10}) admits the first integral
${\cal H}(X,Y,Z):=Z^2 \Omega(X/Z , Y/Z)=h$.
 So, taking into account this first integral we can reduce the dimension
of equation (\ref{biVI10}) and we obtain the following differential equation
\begin{equation}
\frac{d X}{d Z}=\frac{4 h+4 \left(1+k^2\right) X^2+3 \left(1+k^2\right)
(\gamma-2) X Z-4 Z^2}{\left(1+k^2\right) \left[ (2-3 \gamma) h-4 \left(1+k^2\right) X^2
+ (3 \gamma-2) Z^2 \right]},
\label{biVI11}
\end{equation}
which has the particular solution ${\cal F}(X,Z):=-h-\left(1+k^2\right) X^2+Z^2=0$.
The study of differential equation (\ref{biVI11}) allows
us to obtain other integrable Bianchi~VI models. Thus, for instance, if relation
\begin{equation}
1+k^2= \left( \frac{4}{2-3 \gamma} \right)^2,
\label{biVI12}
\end{equation}
is verified, then the straight line
$F_1(X,Z):=4 X+(3 \gamma -2) Z=0$
is an algebraic invariant curve of equation (\ref{biVI11}).
Moreover, turning to the affine plane we obtain that $F_1( \Sigma_+ ,1)=0$
is an invariant straight line of the Bianchi VI models
(\ref{biVI1}) with constraint (\ref{biVI12}).
Now, with this new particular solution we can construct the integrating factor
\begin{equation}
R(\Sigma_+ , N_\times):= N_\times^{\frac{2(4-3 \gamma)}{3 \gamma -2}}
 \Omega^{\frac{9 \gamma -2}{2 (3 \gamma -2)}} F_1( \Sigma_+ ,1).
\label{biVI13}
\end{equation}
The first integral associated to the above integrating factor $R$
is given directly by $\int (q+ 2 \Sigma_+) N_\times R(\Sigma_+ ,
N_\times) \, d  \Sigma_+$. A more detailed study of this first
integral shows that there exists another quadratic invariant
algebraic curve $F_2$ which is functionally independent
of~$\Omega$. In short, we have that
\[
F_2(\Sigma_+ , N_\times):= \frac{(3 \gamma -2)(3 \gamma +2)}{4}
\Sigma_+ + (2+ 3 \gamma) \Sigma_+^2 + (3 \gamma -2) N_\times^2 = 0,
\]
is a new invariant algebraic curve for Bianchi VI
models with constraint (\ref{biVI12}) which allows us to
obtain the following Darboux first integral
\begin{equation}
H(\Sigma_+ , N_\times) = N_\times^{6 (\gamma-2)} \Omega^{2+3 \gamma}
F_2^{2 (2-3 \gamma)}.
\label{biVI14}
\end{equation}

\section{Canonical form of Bianchi VI models}

If we rescale the variable $X$ by ${\cal X}= X \sqrt{1+k^2}$ and
time by $d t = 3 (\gamma - 2)\left(1+k^2\right)dT$, we write
differential equation (\ref{biVI11}) in the following canonical
form
\begin{equation}
\ba{l}
\ds {\cal X}'  =  {\cal X} Z + \lambda_1 {\cal F}( {\cal X} , Z),
\vspace{2mm}\\
Z '  =  {\cal X}^2 + \lambda_2 {\cal F}( {\cal X} , Z) ,
\ea\label{biVI15}
\end{equation}
where the prime denotes derivative respect to $t$.
Now ${\cal F}( {\cal X} , Z) := -h- {\cal X}^2 + Z^2$ is
an algebraic particular solution of the canonical system (\ref{biVI15})
and the parameters $\lambda_1$ and $\lambda_2$ are given by
\[
\lambda_1=\frac{-4}{3( \gamma -2) \sqrt{1+k^2}}, \qquad
 \lambda_2=\frac{3 \gamma -2}{3 (\gamma -2)}.
\]

The integrable Bianchi VI models which arises from parameter
restrictions (\ref{biVI3}), (\ref{biVI7}) and (\ref{biVI12})
correspond to integrable cases of system (\ref{biVI15})
with $\lambda_1 + \lambda_2= 1/2$,
$\lambda_1 + \lambda_2= -1$ and $\lambda_1 + \lambda_2= 0$ respectively.

For $\lambda_1 + \lambda_2= 1/2$,
the canonical system (\ref{biVI15}) has
two new invariant algebraic solutions,
namely ${\cal F}_1( {\cal X} , Z) := \sqrt{h} + {\cal X} + Z$
and ${\cal F}_2( {\cal X} , Z) := \sqrt{h} - {\cal X} - Z$.
Now, performing the change of variable $U={\cal X}+Z$
the canonical system (\ref{biVI15}) becomes
\begin{equation}
\ba{l}
\ds {\cal X}'  =  {\cal X} ( U- {\cal X}) + \lambda_1 \left(-h + U^2-2 U {\cal X} \right),
\vspace{3mm}\\
\ds U '  = \frac{1}{2} \left( U^2 - h\right) .
\ea \label{biVI16}
\end{equation}
Therefore, by integration of the second equation of (\ref{biVI16}),
we have that the general solution of the $U$ variable is given
by the parametrization
\[
U(t)=\sqrt{h} \frac{1- \exp[\sqrt{h} (t-t_0)]}{1+\exp[\sqrt{h} (t-t_0)]},
\]
where $t_0$ is an integration constant. Moreover, the first equation of system
(\ref{biVI16}) is of Riccati type and in addition it has a particular
solution ${\cal X}_1(t)$ which we known from the invariant algebraic
curve ${\cal F}( {\cal X}_1 , U- {\cal X}_1)=0$.
In short, we have ${\cal X}_1(t)= \left[-h + U^2(t)\right] / 2 U(t)$ or more explicitly
\[
{\cal X}_1(t)= \sqrt{h} \, {\rm cosech} [ \sqrt{h} (t-t_0)],
\]
and hence we can obtain the parametrization ${\cal X}(t)$.
Moreover, undoing the changes of variables and rescaling
of the independent variable we obtain the parametrization $X(T)$, $Y(T)$
and $Z(T)$ of the solutions in the complex projective plane in terms
of the dilogarithm special function
${\rm Li}_2(z):=\int_z^0 t^{-1} \log (1-t) \, dt$.
 But the turn to the affine plane is so complex
because the relationship among the physical time $\tau$
and the projective time $T$ is expressed as $d \, \tau = Z(T) \, dT$
and this integration is difficult.

\subsection*{Acknowledgements}
JC is partially supported
by a DGICYCT grant number PB96-1153.

\label{chavarriga-lastpage}

\end{document}